\documentclass[
,showkeys,amsmath,amssymb, aps,prd]{revtex4-2}

\usepackage{color}
\usepackage{graphicx}
\usepackage{dcolumn}
\usepackage{bm}
\newcommand{\eq}[2]{\begin{align}\label{#1}#2\end{align}}

\newcommand{\nn}{\nonumber}
\renewcommand{\ni}{\noindent}
\newcommand{\pa}{\partial}
\newcommand{\ep}{\epsilon}
\newcommand{\al}{\alpha}

\newcommand{\la}{\lambda}
\newcommand{\vphi}{\varphi}\newcommand{\Th}{\Theta}
\newcommand{\vtheta}{\vartheta}

\begin{document}

\title{Tachyon condensation in a chromomagnetic center-vortex background}

\author{M. Bordag}
 \email{bordag@mail.ru}
\affiliation{Bogoljubov Laboratory of Theoretical Physics, Joint Institute for Nuclear Research, 141980 Dubna, Russia}

\date{\today, filename is: \jobname}

\begin{abstract}
\noindent
The chromomagnetic vacuum of SU(2) gluodynamics is considered in the background of a finite radius flux tube (center-vortex) with homogeneous field inside and zero field outside. In this background there are tachyonic modes. These modes cause an instability. It is assumed that the selfinteraction of these modes stops the creation of gluons and that a condensate will be formed. For constant condensates, the minimum of the effective potential is found on the tree level. In the background of these condensates, all tachyonic modes acquire nonzero, real masses which will result in a real effective potential of this system.

Considering only the tachyonic modes and adding the energy of the background field, the total energy is found to have a minimum at some value of the background field, which depends on the coupling of the initial SU(2) model. For small coupling, this dependence is polynomial in distinction from the Savvidy vacuum where it is exponentially suppressed. The minimum of this energy will deepens with shrinking radius of the flux tube. It can be expected that this process can be stopped by adding quantum effects. Using the high temperature expansion of the effective potential, it can be expected that the symmetry, which is broken by the condensate, will be restored at sufficiently high temperature.

\end{abstract}


\maketitle


\section{\label{T1}Introduction}
Quantum Chromodynamics (QCD) is the quantum field theory which describes the strong interactions. It is renormalizable and, thanks to its asymptotic freedom, it is successful in the high energy region, where a perturbative treatment is possible. In distinction, in the low energy region, one observes infrared problems origination from the masslessness of the gluon fields. Physically, one observes confinement, which is, probably, due to a strong multiparticle interaction of gluons and quarks. In addition, and in distinction from QED, the basic fields of QCD do not correspond to the asymptotic states of the theory.

The confinement of gluons and quarks is the main problem left open in the Standard Model. There were many approaches and attempts to solve it. Especially, lattice calculations give a strong support for the idea of confinement and good suggestions for the responsible field configurations; the most convincing one being the dual superconductor configuration. Another approach rests on the functional renormalization group (FRG) by solving flow equations towards some fixed point, see \cite{eich11-83-045014}, for example. A common feature of these approaches is to look for a condensate of gluons that could solve the infrared problems caused by their masslessness, keeping thereby the gauge invariance.

A completely different approach rests on the observation that the magnetic moment of the gluons, due to their spin one, which is twice that of the electron, overcompensates the lowest Landau level in a chromomagnetic field. In a (homogeneous) field $B$, the one particle energy of a gluon,
\eq{1.1}{E_n = \sqrt{p_z^2+gB(2n+1+2s)},
}
may become imaginary in the lowest state ($n=0$ and $s=-1$). Such a state is called {\it tachyonic}. If considering the effective potential, or equivalently, the first quantum corrections to the classical ground state, one arrives at the formula
\eq{1.2}{ V_{eff} &= \frac{B^2}{2}+\frac{11B^2}{48\pi^2}\ln\frac{gB}{\mu}+i\frac{(gB)^2}{8\pi},
}
where $\mu$ is a normalization constant. As first observed in \cite{savv77-71-133}, there is a minimum at some finite $B$, where $V_{eff}<0$, causing the spontaneous generation of such field and forming a new ground state (chromomagnetic or Savvidy vacuum). The reason behind is the coefficient in front of the logarithm, which is the  first coefficient in the beta function and its sign is that of  asymptotic freedom. However, in \cite{niel78-144-376} the imaginary part in \eqref{1.2} was observed, which makes this vacuum state unstable.

There were many attempts to overcome the instability of the chromomagnetic vacuum state. The first one was the so-called {\it Copenhagen vacuum}, see \cite{niel79-160-380}. It rests on the observation that the instability for its formation needs a certain spatial region of a slowly varying background field, for instance to have $gB<p_3^2$ in \eqref{1.1}. One expected a certain domain structure to be formed. Also, in \cite{flor83_SLAC-pub-3244} attention was payed to the observation that the instability occurs from the quadratic part of the action and that the tachyonic modes have a nonlinear, $\phi^4$-type self interaction which acts repulsive. Another approach starts from a selfdual background. In such background, which necessarily involves also a chromoelectric field, the effective potential has also a minimum like \eqref{1.2}, but without imaginary part. In place, one has an infinite number of zero modes \cite{leut80-96-154}. Also, the formulation is in Euclidean space and returning to Minkowski space, the electric field becomes imaginary.

Recently, \cite{savv23-990-116187} was able to sum up these zero modes. Further, there it was shown, that the electric field may be switched off keeping the imaginary part away. As a result, eq. \eqref{1.2} without imaginary part was obtained. A similar result was obtained in \cite{skal00-576-430}, where the gluon polarization tensor was accounted for in some approximation.

It must be mentioned that the Savvidy vacuum has two more unwanted features. The minimum in \eqref{1.2} appears for $gB=\mu^2\exp\left(\frac{24\pi^2}{11g^2}\right)$, i.e., it is exponentially small in a perturbative region. Furthermore, as shown in \cite{ditt81-100-415}, the symmetry breaking, caused by to the chromomagnetic vacuum state, is not restored at high temperature.

The masslessness of the gluon, which is a necessary feature for  the gauge symmetry, hampers all attempts for perturbative calculations in the low energy region. There are many attempts to introduce a mass. As an example we mention \cite{verc08-660-432}, where a special source term was introduced using the formalism of local composite operators introduced by the authors earlier, together with a chromomagnetic background field. However, removing the source, which acts like a gluon mass, brings the instability back.

A decade ago, in \cite{eich11-83-045014} using the functional renormalization group  approach with a selfdual background field, an effective potential like \eqref{1.2} without imaginary part  was found, however with physically more realistic parameters. In \cite{kond14-89-105013}, using a complex  flow equation, a minimum of the effective potential was found, first without magnetic background field. Switching on the magnetic field, the imaginary part re-appeared.

Quite recently, the idea with a domain structure was put forward in \cite{nede21-103-114021} (and citations therein). The domains are assumed to be filled by a selfdual background. The emerging quasi normal modes are treated beyond one loop and the competition between the energy of the domains and the disorder was considered. By minimizing the overall free energy a finite size for the domains was demonstrated.
However, it must be mentioned that all such attempts were unsatisfactory so far.

An attempt to extend the chromomagnetic vacuum from a homogeneous background to a string-like configuration was undertaken in \cite{diak02-66-096004}.
There, for a cylindrical chromomagnetic background field with several profile functions decreasing at infinity, thus having a finite flux and a finite energy (per unit length in direction of the cylinder), the effective potential  was calculated. Such configurations
show a vacuum energy similar to \eqref{1.2}, i.e., some non-trivial minimum. However, in \cite{diak02-66-096004}, no tachyonic mode was seen, which, however, should be there as shown in \cite{bord03-67-065001}.

In \cite{bord23-55-59}, as a new idea for the instability problem, it was suggested to consider a Higgs mechanism for the unstable mode. This mode, as defined by $n=0$ and $s=-1$ in \eqref{1.1}, is a complex field in 2 dimensions ($x_\al$, $\al=0,3$), where $x_3$ is the direction of the magnetic background field), with a negative mass square,
\eq{1.3}{ m^2=-gB<0.
}
In such state, due to the instability, gluon pairs will be created. These are bosons and will form a condensate of tachyons until the process is stopped by their repulsive self interaction. Technically, the potential in the corresponding Lagrangian has a 'Mexican hat' shape and it is necessary to use the Higgs mechanism by making a shift of the tachyon field and to quantize around the shifted field. This was done in \cite{bord23-55-59} for a homogeneous background in radial gauge. Restricting to the lowest orbital mode ($l=0$), one comes to a model with a single complex field in two dimensions. Application of the second Legendre transform in Hartree approximation (or the CJT formalism) resulted in an effective potential with a minimum, as function of the background field, below zero at perturbative values of the parameters (in distinction from \eqref{1.2}, where the minimum for small coupling is exponentially small). Raising the temperature lifts this minimum until it disappears after a certain critical temperature. This way, the symmetry, which was broken by the condensate, is restored.

The paper \cite{bord23-55-59} has several shortcomings. First of all, a homogeneous background field is not really physical, but an approximation at best. Second, the restriction to the lowest orbital momentum mode needs for a better justification. Finally, within the given approach a phase transition in a two dimensional system was seen which seems to be in  contradiction with the Mermin-Wagner theorem.

In the present paper I will solve some of the mentioned problems. I consider as a background a finite radius chromomagnetic  magnetic flux tube and consider all appearing orbital momentum modes of the tachyonic field. Such background is a special case of a center-vortex background, which is frequently discussed in connection with the confinement problem.

The paper is organized as follows. In the next section, the basic formulas for SU(2) are introduced. In the third section, I define the background and the tachyonic mode and derive the corresponding two dimensional Lagrangian. Section IV is devoted to the tachyon condensate and to finding the minimum of the energy. Afterwards go the conclusions. A technical part is delegated to the appendix.

\ni In this paper I use units with $c=\hbar=1$

\section{\label{T2}Basic formulas}
We consider SU(2) gluodynamics in Euclidean space with the Lagrangian
\eq{2.1}{ {\cal L} &= {\cal L}_{\rm YM}+{\cal L}_{\rm gf}+{\cal L}_{\rm gh}
}
where
\eq{2.2}{{\cal L}_{\rm YM}	&=-\frac14 \left(F_{\mu\nu}^a[A]\right)^2,
\\\nn	{\cal L}_{\rm gf}	&=	-\frac{1}{2\xi}	
				\left(D_\mu^{ab}Q_\mu^b\right)
				\left(D_\nu^{ac}Q_\nu^c\right),
\\\nn	{\cal L}_{\rm gh}	&=	\overline{c}^a \left(
		-D_\mu^{ac}D_\mu^{cb}-gD_\mu^{ac}\ep^{cdb}Q_\mu^d \right)c^b,
}
are the Yang-Mills Lagrangian, the gauge fixing term (in background gauge) and the ghost contribution. The field strengths is
\eq{2.3}{ F_{\mu\nu}^a[A]&=\pa_\mu A^a_\nu-\pa_\nu A^a_\mu+g \ep^{abc}A_\mu^b A_\nu^c.
}
A background field $B_\mu^a$ is introduced by
\eq{2.4}{A_\mu^a = B_\mu^a+Q_\mu^a,
}
where $Q_\mu^a$ is the quantum field. The field strength \eqref{2.3} turns into
\eq{2.5}{F_{\mu\nu}^a[B+Q]&=F_{\mu\nu}^a[B]
	+D_\mu^{ab}Q_\nu^b-D_\nu^{ab}Q_\mu^b
	+g \ep^{abc}Q_\mu^bQ_\nu^c
}
with the covariant derivative
\eq{2.6}{	D_\mu^{ab} &= \pa_\mu\delta^{ab}+g\ep^{acb}B_\mu^c.
}
We mention its commutator,
\eq{2.7}{\big[D_\mu ,D_\nu \big]^{ab} = g \ep^{acb}F_{\mu\nu}^c[B].
}
With the background field $B_\mu^a$, introduced in \eqref{2.4}, the gauge fixing Lagrangian ${\cal L}_{\rm gf}$ in \eqref{2.3} defines an $R_\xi$ gauge. In the following we put $\xi=1$, i.e., we work in the Feynman gauge. We mention that when not going going beyond the one loop approximation in the effective action, the gauge invariance  should be guaranteed.

We insert \eqref{2.5} into \eqref{2.2},
\eq{2.8}{{\cal L}_{\rm YM} &=	
	{\cal L}_0+{\cal L}_1+{\cal L}_2+{\cal L}_3+{\cal L}_4,
}
where
\eq{2.9}{{\cal L}_0	&=	-\frac14 \left(F_{\mu\nu}^a[B]\right)^2
}
is the (classical) background contribution,
\eq{2.10}{ {\cal L}_1	&=	Q_\nu^a D_\mu^{ab} B_{\mu\nu}^b
}
with $B_{\mu\nu}^b =\pa_\mu B_\nu^b-\pa_\nu B_\mu^b$ is the linear term (source term), and the remaining contributions,
\eq{2.11}{{\cal L}_2&=-\frac12 Q_\mu^a \Bigg(
	-(D_\la^a)^2\delta_{\mu\nu}
	-2g \ep^{acb}B_{\mu\nu}^c
	\Bigg) Q_\nu^b,
\nn\\	{\cal L}_3	&=	-g\ep^{abc}(D_{\mu\nu}^{ad}Q_\nu^d)Q_\mu^b Q_\nu^c,
\\\nn {\cal L}_4	&=	-\frac{g^2}{4}(Q_\mu^a Q_\mu^a Q_\nu^b Q_\nu^b
								-	Q_\mu^a Q_\nu^a Q_\mu^b Q_\nu^b ).
}
are quadratic, cubic and quartic in the quantum field.

In the following we consider an Abelian background field
\eq{2.12}{ B_\mu^a=\delta^{a3}B_\mu,
	~~~~~B_{\mu\nu}=\pa_\mu B_\nu-\pa_\nu B_\mu.
}
With this, it is convenient to turn into the so-called {\it charged basis}, which diagonalizes the Lagrangian in color space. Doing the corresponding substitutions,
\eq{2.13}{ Q_\mu^1 	&= \frac{1}{\sqrt{2}} (W_\mu +W^*_\mu),~~~&Q_\mu^3 &=Q_\mu,
\\\nn		 Q_\mu^2 	&= \frac{1}{\sqrt{2i}} (W_\mu -W^*_\mu), & W_\mu&=\frac{1}{\sqrt{2}}\left(Q_\mu^1+i Q_\mu^2\right).
}
in \eqref{2.11}, we arrive at
\eq{2.14}{
{\cal L}_2	&=-\frac12 Q_\mu  \Big(
	-\pa^2\delta_{\mu\nu}
	\Big) Q_\nu
\\\nn&~~~~~~~~~~~~~~
	- W^*_\mu \Big(
	-(D_\la)^2\delta_{\mu\nu}
	-2ig B_{\mu\nu} 	\Big) W_\nu,
\\\nn	
{\cal L}_3	&=	-ig(Q_\mu W^*_{\mu\nu}W_\nu-Q_\mu 	
	W_{\mu\nu}W_\nu^*-Q_{\mu\nu}W^*_\mu W_\nu)
	\\\nn
{\cal L}_4	&=	-{g^2}(
	 Q_\mu Q_\mu W^*_\nu W_\nu
	-Q_\mu Q_\nu W^*_\mu W_\nu
\\\nn&~~~~~~~~~~~~~~	+W_\mu^* W_\mu W^*_\nu W_\nu
	-W_\mu^* W_\nu W^*_\mu W_\nu).
}
The third component, $Q_\mu$, is interpreted as color neutral vector field  whereas $W_\mu$ represents a  color charged vector field.
In \eqref{2.14} we introduced the notations
\eq{2.15}{ Q_{\mu\nu}=\pa_\mu Q_\nu-\pa_\nu Q_\mu,~~
	W_{\mu\nu}=D_\mu W_\nu-D_\nu W_\mu,
}
and the covariant derivative for the charged field is
\eq{2.16}{ D_\mu=\pa_\mu-i B_\mu,\quad [D_\mu,D_\nu]=-iB_{\mu\nu}.
}
In the following section we specialize these general formulas to the case of a cylindrical symmetric background field.

\section{\label{T3}Cylindrically symmetric background field and a field theory for the tachyonic mode}
We consider a cylindrical symmetric chromomagnetic magnetic background field, for instance a straight vortex line, parallel to the third spatial axis.  In cylindrical coordinates $(r,\varphi,x_3)$ we take the upper two components of the  potential $B_\mu$, $\mu=1,2$, in the form (in two dimensional vector notations)
\eq{3.1}{ \vec B=\vec e_\varphi\frac{\mu(r)}{r},
}
together with $B_3=B_4=0$. The radial profile $\rho(r)$ is, for the moment, arbitrary. The field strengths in \eqref{2.12} turnes into
\eq{3.2}{B_{\mu\nu}=\ep_{\mu\nu}\frac{\mu'(r)}{r},
	~~~~\ep_{\mu\nu}=\left(\begin{array}{cccc}
		0&1&0&0\\-1&0&0&0\\
		0&0&0&0\\0&0&0&0   \end{array}\right)_{\mu\nu}
}
and for the commutator we obtain
\eq{3.2a}{[D_1,D_2]=-i \, \frac{\mu'(r)}{r}.
}
We mention that in our notations $B_\mu$ in \eqref{2.11} is the potential and $B_{\mu\nu}$ is the field strength of the background field. In \eqref{3.2},
\eq{3.3}{\frac{\mu'(r)}{r}\equiv B(r),
}
has the meaning of the modulus of the three dimensional field strength belonging to the vector potential \eqref{3.1}.

Next we consider the linear term \eqref{2.10}. With \eqref{2.6} and \eqref{2.10} we get
\eq{3.4}{{\cal L}_{1} =Q_\nu^3\pa_\mu\ep_{\mu\nu}\frac{\mu'(r)}{r}
}
and using \eqref{3.2} we arrive at
\eq{3.5}{{\cal L}_{1} = (-\sin\varphi Q^3_1+\cos\varphi Q^3_2)\left(\frac{\mu'(r)}{r}\right)'.
}
In a homogeneous background field, where $B(r)=const$ in \eqref{3.3}, ${\cal L}_{1}$ vanishes. In a non-homogeneous background, which we will   consider below, it does not vanish. However, it couples to the third color component, i.e. to the color neutral one, which does not influence the tachyonic mode.

Next we define the tachyonic mode. We consider the spectrum of the operator representing the kernel of the quadratic part in $W_\mu$, ${\cal L}_2$ in \eqref{2.14}. The corresponding wave equation reads
\eq{3.6}{ \left((D_\la)^2\delta_{\mu\nu}-2igB_{\mu\nu}\right)W_\nu=0.
}
In a homogeneous background we have a profile function
\eq{3.7}{ \mu(r)=\frac{Br^2}{2},~~\frac{\mu'(r)}{r}=B,
}
with a constant $B$ and, after Fourier transform in time and $z$-directions, the spectrum is well known,
\eq{3.8}{ k_0^2=k_3^2+gB(2n+1+2s),
}
where $n=0,1,\dots$ enumerates the Landau levels and $s=\pm1$ is the spin projection. The tachyonic mode is defined by $n=0$, $s=-1$ and its spectrum,
\eq{3.9}{ k_0^2=k_3^2-gB,
}
has a negative eigenvalue, which can be interpreted as a negative mass square, $m^2=-gB$. As mentioned in the Introduction, this is the reason to call it tachyonic. Also, frequently it is called {\it the unstable mode}.
In fact, this is not a single mode. In the eigenvalue problem \eqref{3.6}, there is a further quantum number, the orbital momentum, with respect to which the spectrum is degenerated. This corresponds to the translational invariance of the problem in the plane perpendicular to the magnetic field. In this sense, there are infinitely many tachyonic modes.

\begin{figure}[h]  
	\includegraphics[width=0.5\textwidth]{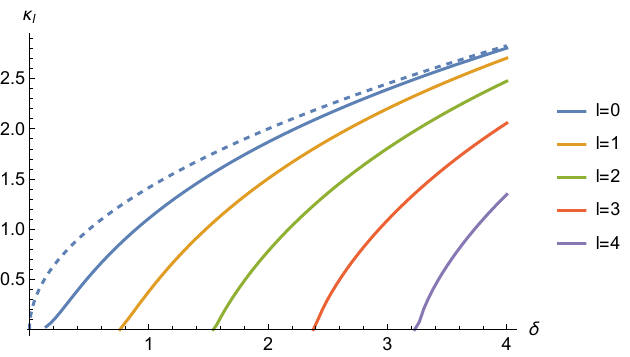}	
		\includegraphics[width=0.4\textwidth]{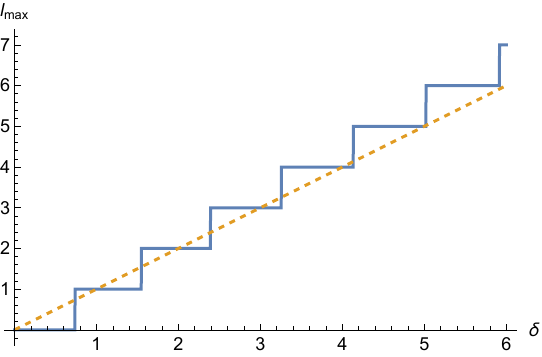}	 
		\caption{Left panel: The tachyonic levels \eqref{3.12} in the flux tube \eqref{3.13}.
	Right panel: The maximal number $l_{max}$ of orbital momenta for a given flux $\delta$ (solid line). This is the number of curves crossed by a vertical section in the left panel. For comparison, the dashed line shows $\delta$.
	}\label{fig:1}
\end{figure}

In the present paper we consider a magnetic field which is homogeneous inside a cylinder of radius $R$ and zero outside. In this case the profile function is
\eq{3.10}{ \mu(r)=\frac{Br^2}{2}\Theta(R-r)+\frac{BR^2}{2}\Theta(r-R),~~\frac{\mu'(r)}{r}\equiv B(r)=B \, \Theta(R-r),
}
This field has a finite flux, $\Phi$, and a finite energy, $E_{bg}$,
\eq{3.11}{ \Phi=\int d^2x_\perp  B(r)=\pi R^2B,
	~~~E_{bg}=\frac12\int d^2x_\perp B(r)^2=\frac{\pi}{2}B^2R^2,
}
the energy being a density per unit lengths of the third direction.
In this background, the spectrum of the color charged field $W_\mu$, \eqref{2.13}, is more complicated then \eqref{3.9}. Nevertheless it has tachyonic modes,
\eq{3.12}{ k_0^2=k_3^3-\kappa_l^2,
}
with $\kappa_l^2>0$, see Figure \ref{fig:1}.

In this background, for the tachyonic modes $ W^{\rm ta}_\mu(x)$ we consider the mode decomposition
\eq{3.13}{ W^{\rm ta}_\mu(x)
	=\frac{1}{\sqrt{2}}
	\left(\begin{array}{c}1\\ i \\ 0\\ 0\end{array}\right)_\mu
	\sum_{l=0}^{l_{max}} \frac{e^{i l\varphi}}{\sqrt{2\pi}}\phi_l(r) \psi_l(x_\al),~~~~(\al=0,3)
	~~~x_\perp=(r,\varphi),
}
($x_\perp$ taken in cylindrical coordinates), $l_{max}$ is discussed below. In \eqref{3.13}, the $\kappa_l$ are the eigenvalues \eqref{3.14} and $ \phi_l(r)$ are the eigenfunctions of the spatial part of the operator in \eqref{3.6},
\eq{3.14}{  \left(
	 \pa_r^2+\frac{1}{r}\pa_r-\frac{(l-\mu(r))^2}{r^2}
 +2\frac{\mu'(r)}{r} \right) \phi_l(r)=\kappa_l^2 \phi_l(r).
}
where, with \eqref{3.2},
\eq{3.15}{\ep_{\mu\nu}\left(\begin{array}{c}1\\ i \\ 0\\ 0\end{array}\right)_\nu=i\left(\begin{array}{c}1\\ i \\ 0\\ 0\end{array}\right)_\mu
}
was used. In the mode decomposition \eqref{3.13}, the coefficients
$\psi_l(x_\al)$ are the free coefficients which in the procedure of canonical quantization become the operators. These $\psi_l(x_\al)$ are complex fields  depending on two variables, $x_3$ and $x_0$ (or $x_4$ in an Euclidean version).

The  eigenvalue problem \eqref{3.14}, describing the tachyonic modes \eqref{3.13}, has scattering solutions and bound state solutions as well. The scattering solutions have $\kappa_l=ik$ and, of course, a continuous spectrum. In opposite, the bound state solutions have real $\kappa_l$, are normalizable  and have a discrete spectrum. In the following we consider only these solutions.
The methods for solving the eigenvalue problem \eqref{3.14} are well known. We demonstrate their application in the Appendix. There are no analytical formulas, but the numerical evaluation is quite easy using standard methods. We demonstrate the result in figure \ref{fig:1} (left panel) as function of
\eq{3.21}{\delta=\frac{BR^2}{2},
}
which by means of $\delta=\Phi/2\pi$, is related to the magnetic flux \eqref{3.11}. We mention that for $\delta<0.08$ there is no solution, which is similar to the restriction $gB\le p_3$, now in the transversal direction. Increasing the flux, with each new flux quantum, one new solutions comes down from the continuum. In the limit of $R\to\infty$, one will see all the degenerated solutions known from the homogeneous field. In figure \ref{fig:1}, in this limit all curves will merge into the dashed line. Spelled out in the reverse order, the finite extend of the magnetic field splits the tachyonic levels, i.e., it removes the degeneracy,  and there is now a finite number of them. In figure \ref{fig:1}, the dependence of the maximal number $l_{max}$ of orbital momenta for a given flux $\delta$ is shown. For large flux, $l_{max}\to\delta$ holds.

\begin{figure}[h]  
	\includegraphics[width=0.45\textwidth]{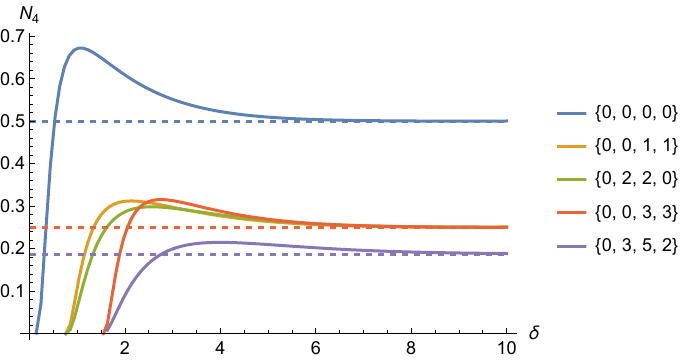}	 
	\caption{ The coefficients $N_4(l_i)$ for sets  $\{l_1,l_2,l_3,l_4\}$, from top to bottom,  $\{0,0,0,0\}$,$\{0,0,1,1\}$,$\{0,2,2,0\}$,\dots\,. The dashed lines show the corresponding quantity in the case of a homogeneous background.
	}\label{fig:0}
\end{figure}

The next step is to set up a field theory for the tachyonic mode. With eqs. \eqref{3.13} and \eqref{3.16} we have a mode expansion for the tachyonic modes. Similar expansions could be set up for all other, non-tachyonic  modes too. However, as said above we restrict yourself to the tachyonic modes. The reason is that   the other modes are stable. 
 A theory with only the tachyonic modes appears when inserting \eqref{3.13} into \eqref{2.14} and integrating over the transversal coordinates. We obtain
\eq{3.28a}{ {\cal \tilde L}=\int dx_\perp \ {\cal L}\equiv {\cal \tilde L}_2+{\cal\tilde L}_4
}
with
\eq{3.28}{{\cal\tilde L}_2 &
	=-\sum_{l=0}^{l_{max}} \psi^*_l(x_\al)\left(-\pa_\al^2+m_l^2\right)\psi_l(x_\al),
\\\nn
	{\cal\tilde L}_4 & =-\la \delta \sum_{l_1,\dots,l_4\le l_{max}}\delta_{l_1-l_2,l_3-l_4}N_4(l_i)\
	\psi^*_{l_1}(x_\al)\psi_{l_2}(x_\al)\psi^*_{l_3}(x_\al)\psi_{l_4}(x_\al)
}
and the coefficients are
\eq{3.29}{ m_l^2&=-\kappa_l^2,~~~~\la=\frac{g^2}{\pi},
\\\nn
	N_4(l_i)&=\int_0^\infty  dr\, r\,
	\phi_{l_1}(r)\phi_{l_2}(r)\phi_{l_3}(r)\phi_{l_4}(r).
}
All these quantities depend on the flux $\delta$ as a parameter. In these formulas we have put $R=1$. The dependence on $R$ can be restored simply by dividing \eqref{3.28a} by $R^2$.

The masses $m_l$ are the (imaginary) masses of the tachyonic modes.
The coefficients $N_4(l_i)$ in the quartic contribution depend on the flux $\delta$ and can be calculated by inserting \eqref{3.16} into the lower line in \eqref{3.29}. We mention that these are symmetric in the arguments $l_i$. Some examples are shown in figure \ref{fig:0}. 
The case of a homogeneous background can be obtained from $R\to\infty$ in \eqref{3.16} and \eqref{3.29}. In that case the wave functions and the integration are explicit and result in
\eq{3.30}{N_4^{hom}(l_i) &= \frac{\Gamma(l_1+l_3+1)}{2^{l_1+l_3}
		\sqrt{\Gamma(l_1+1)\Gamma(l_2+1)\Gamma(l_3+1)\Gamma(l_4+1)}}.
}
These are the dashed lines in figure \ref{fig:0}. In the limit $R\to\infty$, as can be seen in  the Appendix, simply the exterior solution \eqref{3.17} is exponentially small and the limit is reached   exponentially fast. This is also the speed with which in the figure the dashed lines are reached. However, since for a given $l$, some solutions start at $\sim\delta$, there are for all $\delta$ contributions which are far from the homogeneous limit \eqref{3.30}.

The Lagrangian \eqref{3.28a} describes a theory with a finite number of complex fields in two dimensions with negative mass square, \eqref{3.29}, and a dimensional coupling ($\la$, \eqref{3.29}, is dimensionless and was introduced for convenience). The eigenvalues $\kappa_l^2$ in \eqref{3.29} and the factor $	N_4(l_i)$ have dimension $R^{-2}$ (when restoring $R$) and their magnitude depends on the magnetic background field $B(r)$, resp. using the specific background \eqref{3.10}, on $B$.
We remind, that there is also the classical background \eqref{2.9} with the energy \eqref{3.11}.

\section{\label{T4}A stable tachyon condensate}
%
As discussed in \cite{bord2207.08711}, I assume that the tachyonic modes will form a condensate similar to the scalar field in the well known Higgs model. With \eqref{3.28a}, we have a system with a kind of {\it Mexican hat potential} for the fields $\psi_l(x_\al )$. The negative mass square in ${\cal\tilde L}_2$ makes the system for $\psi_l(x_\al)=0$ 'sitting on the top of the hill' and causes the effective potential to have an imaginary part, which is just that which was observed in \cite{niel78-144-376} for a homogeneous background. In the preceding section we have seen that these instabilities appear in the inhomogeneous background \eqref{3.10} too. The imaginary part makes the system unstable and pushes it forwards to a state with lower energy until the imaginary part disappears. There are, probably, many ways for QCD to go to a lower state. Here we consider those, which are within the model defined by the Lagrangian \eqref{3.28a}. Thus the system will create modes of the field $\psi_{l_1}(x_\al )$, i.e., tachyons, until it is stopped by the repulsive self-interaction given by ${\cal \tilde L}_4$ in \eqref{3.28}. These modes will form a Bose condensate. The situation is similar to a quartic oscillator in quantum mechanics with an imaginary frequency and a $x^4$-term entering with a plus sign. So we have to look for the minimum of the potential $-\cal\tilde  L$. We mention that the existence of a minimum is guaranteed by the structure of $\cal\tilde L$ since all coefficients in ${\cal\tilde L}_4$, \eqref{3.28}, are positive.

In \cite{bord2207.08711}, only one orbital momentum mode, $l=0$, was allowed and the above idea was realized by a shift, $\psi (x_\al )\to \psi(x_\al )+v$, of this mode, where $v$ is a constant condensate. In the present case we allow for all orbital momentum modes in \eqref{3.13} and we have a correspondingly more complicated situation. As said above, we have to consider the minimum of $-{\cal\tilde L}$, \eqref{3.28a}.
In view of a later quantization, this means that  we consider a minimum of the effective potential (which would include quantum corrections) on the tree level.

We parameterize the complex fields,
\eq{4.1}{ \psi_l(x_\al) &= \frac{1}{\sqrt{2}}\vphi_l(x_\al)\, e^{i\Th_l(x_\al)}
}
by two real fields,  $\vphi_l(x_\al)$ and $\Th_l(x_\al)$, having a meaning of module and phase. We mention that ${\cal\tilde L}_4$ in \eqref{3.28} is real, which is ensured by the Kronecker symbol in \eqref{3.28}, i.e., by the orbital momentum conservation.

In the following, we look for a minimum on constant fields. An inhomogeneity only tends to increase  the energy. Of course, a minimum on non-constant fields cannot be excluded in such simple way. However, this is a separate problem and left for later.  Since there is no way to get analytical results, we are left with numerical methods. The calculations were preformed by Mathematica, using the tools provided by that system.

To look for a minimum we make shifts of the fields,
\eq{4.2a}{\vphi_l(x_\al) \to v_l +  \vphi_l(x_\al),~~~
	\Theta_l(x_\al)\to \vtheta_l+\Theta_l(x_\al)
}
with constant $v_l$ and $\vtheta_l$.

We insert \eqref{4.1} and \eqref{4.2a} into the Lagrangian \eqref{3.28a} and expand for small $\vphi_l(x_\al)$ and $\Th_l(x_\al)$. We arrive at
\eq{4.3}{{\cal \hat L}&= {\cal \hat L}_0+{\cal \hat L}_1+{\cal \hat L}_2+\dots
}
with
\eq{4.4}{ {\cal \hat L}_0 &= \frac12 \sum_{l=0}^{l_{max}}   \kappa_l^2v_l^2
	-\frac{g^2}{2\pi} \delta \sum_{l_1,\dots,l_4\le l_{max}}\delta_{l_1-l_2,l_3-l_4}N_4(l_i)\
	v_{l_1}v_{l_2}v_{l_3}v_{l_4}\ e^{i\vtheta_{l1}-i\vtheta_{l2}+i\vtheta_{l3}-i\vtheta_{l4}},
\\\nn
	{\cal \hat L}_1 &=   \sum_{l=0}^{l_{max}} \left[  \kappa_l^2v_l
	-4\frac{g^2}{2\pi} \delta \sum_{l_1,\dots,l_4\le l_{max}}\delta_{l_1-l_2,l_3-l_4}N_4(l_i)\
 	\delta_{l,l_1}v_{l_2}v_{l_3}v_{l_4}\right]\vphi_l(x_\al),
\\\nn
{\cal \hat L}_2 &=  -\sum_{l=0}^{l_{max}}
	\big[	\frac12\vphi_l(x_\al)\left(-\pa_\al^2\delta_{ll'}
						+m^2_{ll'}\right)\vphi_l(x_\al)
						+v_l^2\Th_l(x_\al)(-\pa_\al^2)\Th_l(x_\al)						
		\big].
}
The mass is now a matrix with entries
\eq{4.5}{m_{ll'}^2 &= -\kappa_l^2 \delta_{ll'}
	+3\frac{g^2}{2\pi} \delta \sum_{l_1,\dots,l_4\le l_{max}}
	\delta_{l_1-l_2,l_3-l_4}N_4(l_i)\delta_{l,l_1}
	\delta_{l,l_2} \  v_{l_3}v_{l_4}.
}
We are looking for a minimum of $ {\cal \hat L}_0$, which at once is a zero of $ {\cal \hat L}_1$. We mention that  the $\vtheta_l(x_\al)$ do not cancel in ${\cal \hat L}_0$. The first appearance of a $\vtheta_l$ is for $\delta=3.03$,  where we have orbital momenta until $l=3$,
%
%
\eq{4.6}{{\cal\hat L}_0&=-2.89926 v_0^2-2.46514 v_1^2-1.67909 v_2^2-0.640937 v_3^2
		\\\nn  &
	+\lambda  \big[
	0.218062 v_1^2 v_2 v_0 \cos \left(\vtheta_0-2 \vtheta_1+\vtheta_2\right)
	+0.256459 v_1 v_2 v_3 v_0 \cos
	\left(\vtheta_0-\vtheta_1-\vtheta_2+\vtheta_3\right)
	\\\nn&
	+0.158381 v_1 v_2^2 v_3 \cos \left(\vtheta_1-2
	\vtheta_2+\vtheta_3\right)
	+0.137526 v_0^4+0.296608 v_1^2 v_0^2+0.161249 v_2^2 v_0^2
	+0.0745899 v_3^2 v_0^2
	\\\nn&
	+0.0737236
	v_1^4+0.0502424 v_2^4
	+0.0234354 v_3^4
	+0.220456 v_1^2 v_2^2
	+0.124823 v_1^2 v_3^2
	+0.130519 v_2^2 v_3^2             \big],
}
to show an example.

It is to be mentioned that the minimum may be not unique.
In the example \eqref{4.6}, a change in the angles $\vtheta_i$, keeping the arguments of the cosines, is possible since that would imply 3 conditions for 4 variables. In the following we consider only one minimum. As it turned out, one of them is realized for all $\vtheta_l=0$. For this reason we dropped the  $\vtheta_l$ in ${\cal \hat L}_1 $. In \eqref{4.4}, the last line is the quadratic part of the Lagrangian. It is not non-diagonal in the fields $\vphi_l(x_\al)$.  The fields $\Th_l(x_\al)$ remain massless and in the sense of a spontaneous symmetry breaking these are the Goldstone bosons. In this spirit, we call the    $v_l$, which realize the minimum of ${\cal \hat L}_1 $,  the {\it tree level condensates}, $v^{tree}_l$ and the value of ${\cal \hat L}_0$,
\eq{4.7}{V_{eff}^{tree}\equiv -{{{\cal\hat L}_0}} {}_{\big | {v_l^{tree}}}\ ,
}
the effective potential on tree level.

\begin{figure}[h]  
	\includegraphics[width=0.4\textwidth]{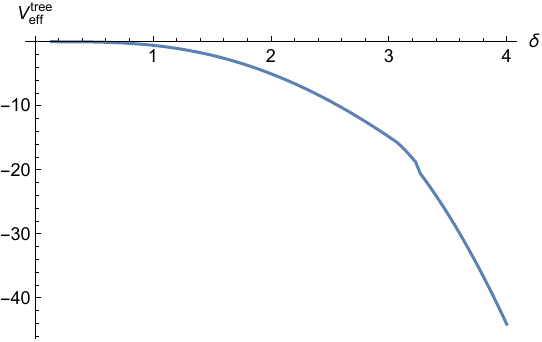}	 
	\includegraphics[width=0.5\textwidth]{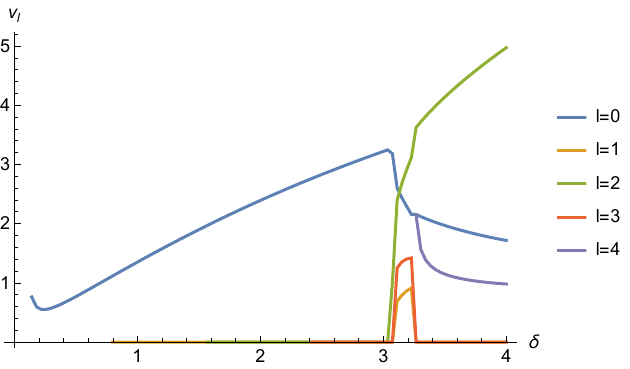} 
	\caption{Left panel: The value of $V_{eff}^{tree}=-{\cal L}$, \eqref{3.28a} in the minimum.
		Right panel: The tree level condensates $v^{tree}_l$ as function of the flux $\delta$. The mesh for these plots is $\Delta\delta=0.039$.
	}\label{fig:2}
\end{figure}

Some of the first (in the sense of increasing flux $\delta$)  condensates and $V_{eff}^{tree}$ are shown in figure \ref{fig:2} (right panel) for   $\delta\le4$. The depth of the minimum of $V_{eff}^{tree}$ grows with the flux. Until $\delta=3.03$ we have only one non-zero condensate, $v^{tree}_1$, beyond all components may be non-zero.

The behavior of the condensates deserves special attention. Until $\delta=3.03$, we have up to 4 orbital momentum modes present (see figure \ref{fig:1}) and only one non zero condensate. At $\delta=3.03$, without exciting a new orbital mode, the behavior changes drastically; now all condensates $v_l$ ($l=0,...,3$)   are non zero, see  figure \ref{fig:2}, right panel. At $\delta=3.25$, the mode $l=4$ sets in and here we have each second condensate nonzero. We did not investigate this behavior in more detail, but we assume that here classical chaos can be observed.

In the minimum of $V_{eff}^{tree}$, the first order variation in $\vphi_l(x_\al)$, i.e., $ {\cal \hat L}_1$, vanishes. This circumstance was used in the numerical calculations as a check for the procedure to find the minimum. For instance, the expression
\eq{4.8}{   \sum_{l=0}^{l_m} \big|
	  -\kappa_l^2(v_l^{tree})^2
	+\la \delta \sum_{l_i}\delta_{l,l_1}N_4(l_i) \delta_{l_1-l_2,l_3-l_4}
	v_{l_1}^{tree}v_{l_2}^{tree}v_{l_3}^{tree}v_{l_4}^{tree} \big |,
}
which accumulates the mismatches from the first variations, was seen to be below $10^{-9}$ for all calculated values of $\delta$. In order to reach this, in the integration in \eqref{3.29}, in the routines for finding the minima of $-\cal L$ and for solving eq. \eqref{3.24} for $\kappa_l$, a working precision of 100 digits was used.

\begin{figure}[h]  
	\includegraphics[width=0.5\textwidth]{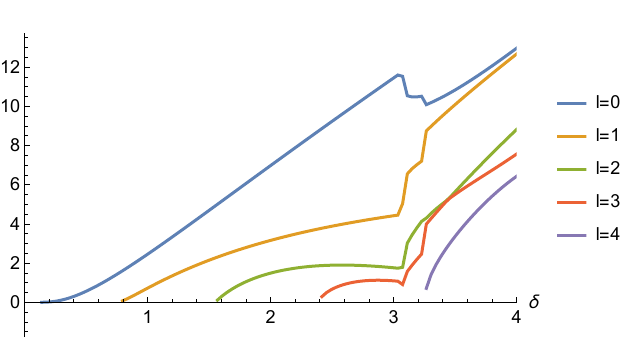}	
	\caption{The mass eigenvalues on tree level, i.e., after diagonalization of \eqref{4.6}.
	}\label{fig:3}
\end{figure}

\begin{figure}[h]  
	\includegraphics[width=0.48\textwidth]{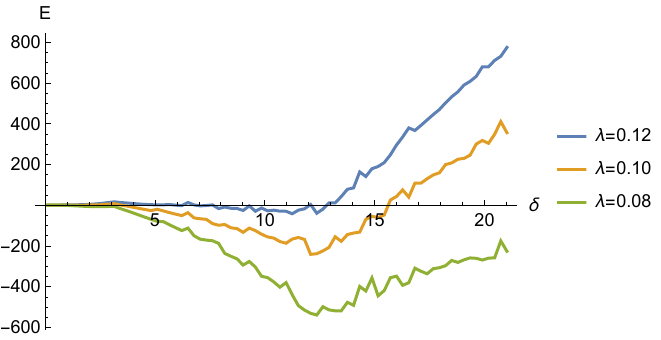}	
	\includegraphics[width=0.48\textwidth]{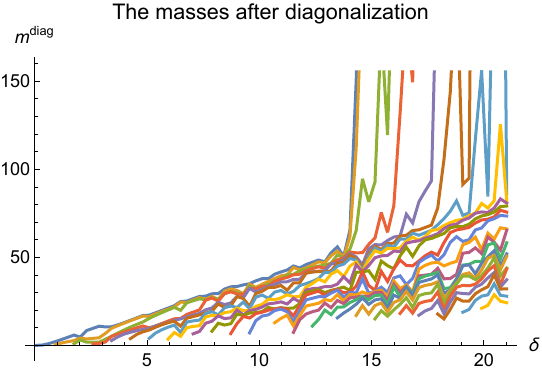}	
	\caption{Left panel: The energy $E$, \eqref{4.7}, of the system. Right panel: The mass eigenvalues on tree level, i.e., after diagonalization of \eqref{4.6}, for $\la=0.1$.  The mesh for these plots is $\Delta\delta=0.28$.
	}\label{fig:4}
\end{figure}

In the second order variation, $ {\cal \hat L}_2$, we have a mass matrix, \eqref{4.5}. It can be diagonalized and the eigenvalues are shown in figure \ref{fig:3}  (and in figure \ref{fig:4} for larger $\delta$). These are all non negative and grow with the flux.

As can be also observed from these figures, after $\delta=3.03$, the behavior becomes a bit irregular (similar to that of the condensates in figure \ref{fig:2}), keeping however its basic features. The same holds for the energy in figure \ref{fig:4} (left panel). The unevenness  of the curve are not due to numerical errors, but are an intrinsic property.

The minimum of the complete energy, i.e., with the energy \eqref{2.9}, or \eqref{3.11}, of the background field added, is
\eq{4.9}{ E&=E_{bg}+V_{eff}^{tree}
}
with $\cal L$ from \eqref{3.28a} with $v_l^{tree}$ inserted. Restoring the $R$-dependence, a factor $R^{-2}$ must be added.  As a function of $\delta$, the energy $E$ is shown in figure \ref{fig:4} (left panel) for several values of the coupling $\la$. As can be seen, for large $\la$ there is no minimum. It appears at $\la\sim0.12$ and deepens with decreasing $\la$. In the right panel of figure \ref{fig:4}, $m^{diag}$, i.e., the eigenvalues of the mass matrix \eqref{4.5}, are shown. This is a continuation of figure \ref{fig:3} to larger $\delta$. All these masses are positive, some become large.

\begin{figure}[h]  
	\includegraphics[width=0.48\textwidth]{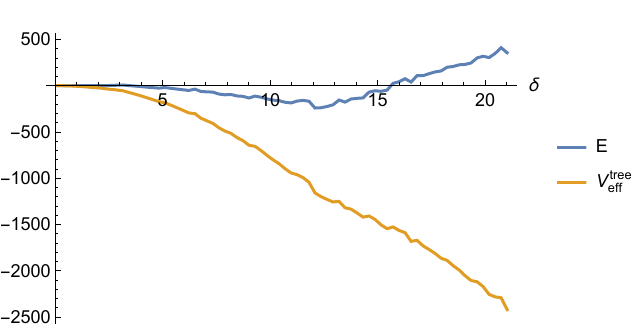}	
	\includegraphics[width=0.48\textwidth]{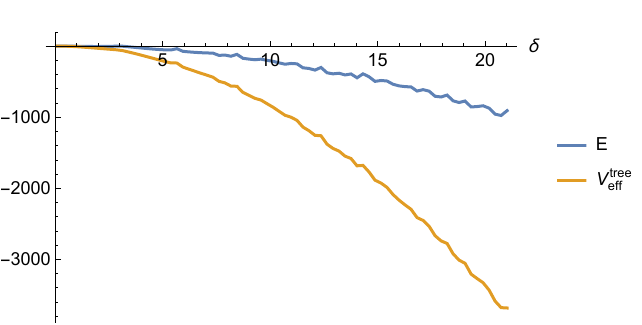}	
	\caption{Left panel: The effective potential  \eqref{4.7} and the energy \eqref{4.9} for $\la=0.1$. Right panel:  The effective potential  \eqref{4.7} and the energy \eqref{4.9} for $\la=0.1$ calculated with $N_4^{hom}$, \eqref{3.30}, in place of $N_4$, \eqref{3.29}.
	}\label{fig:5}
\end{figure}

It is interesting to mention that the effective potential \eqref{4.7}, which is shown in figure \ref{fig:2} for small $\delta$, continues to grow in negative direction also for larger $\delta$, as shown in figure \ref{fig:5} in comparison with the energy $E$, \eqref{4.9}.

Now let us discuss the relation to a homogeneous magnetic background. Formally it corresponds to an infinite radius in our model, $R\to\infty$. Of course the energy diverges as being proportional to the area in the directions perpendicular to the magnetic field. Equivalently, in the radial gauge, the number of orbital momenta involved diverges. Therefor a regularization is needed. As such just the finite radius, considered in this paper, may be taken. For instance, it provides a restriction for the orbital momenta, $l_{max}$, see figure \ref{fig:1}. Considering $R$, or  $l_{max}$, as regularization, we have the same formulas as before with the only change that we may take $N_4^{hom}$, \eqref{3.30}, in place of $N_4$, \eqref{3.29}. The result is shown in figure \ref{fig:5}, right panel. Both, the effective potential and the energy are below the corresponding values in the left panel, and, in addition, the energy has non minimum. This behavior demonstrates for instance that taking a restriction of the angular momenta as regularization for a calculation in the homogeneous background, gives wrong results.

\section{\label{T5}Discussion and conclusions}
In the preceding section we have seen that in a string like chromomagnetic background the tachyonic modes of the gluon field will form a condensate. We took for the background a homogeneous field inside a cylinder of radius $R$ and zero field outside. It has a finite flux $\Phi$ and a finite energy $E_{bg}$ (per unit length of the cylinder), \eqref{3.11}. Due to the cylindrical symmetry, we have orbital modes of the tachyon field, $\psi_l(a_\al)$, \eqref{3.13}. Their number is restricted by the flux, $0\le l\lesssim\delta$ (see figure \ref{fig:1}, right panel). Each orbital momentum mode   may have a condensate, whereby we considered only constant condensates of the module $\vphi_l(a_\al)$, \eqref{4.1}.

To find the minimum of $-{\cal L}_0$, \eqref{4.4}, is a task in several variables. We used the numerical capabilities provided by Mathematica. The depth of the emerging minimum is shown in  figure \ref{fig:2} (left panel) and, for larger $\delta$ in  figure \ref{fig:5} (also left panel, upper curve), as function of the flux $\delta$. It takes negative values and grows with the flux.

An unexpected feature is the structure of the minima for $\delta>3.03$. As can be seen in  figure \ref{fig:2} (right panel), and also in  figure \ref{fig:3}, the behavior of the solutions changes drastically, while, however keeping the basic features. We interpret this phenomenon as the onset of classical chaos.

The minimum $V_{min}$ of the effective potential $V_{eff}^{tree}$, \eqref{4.7},   deepens with growing flux. To get the total energy, one has to add the energy of the background field $E_{bg}$, \eqref{3.11}. Using \eqref{3.21} one comes to
\eqref{4.9}. Restoring the $R$-dependence, it can be written in the form
\eq{5.1}{ E=\frac{\pi \delta+V_{min}}{R^2}.
}
As can be seen from  figure \ref{fig:4} (left panel), for a coupling $\la\lesssim 0.12$, it has a minimum at some finite $\delta$. With fixed radius $R$, the corresponding value of the magnetic field would be chosen by the system automatically.

In case, one allows also the radius being a dynamical variable, the system would   prefer $R\to0$ as the direction lowering the energy. We mention that this is a result of our purely classical consideration.  We may hope that this shrinking of the radius will be stopped when quantum effects are included.

We started from a SU(2) chromodynamics. We separated the tachyonic modes and have seen that these will create a chromomagnetic background field and will form a stable condensate. It is to be mentioned that this approach is in distinction from the most common assumption of the condensate for all gluon modes which is motivated by keeping the gauge symmetry. An attractive feature of our approach is that  a condensate of the tachyon modes provides masses to all these modes, see  figure \ref{fig:4} (right panel). This is similar to the Higgs mechanism in the Standard model.

A task for further investigation of this approach is the calculation of the vacuum energy of the tachyonic modes. Since all these have  real, non zero masses (see  figure \ref{fig:4} (right panel), this should not be a problem. Moreover, when including temperature, a simple estimation for high $T$, following eq. (49) in \cite{bord23-55-59}, shows that additional contributions $\sim \sum_l m_lT$, can be expected, removing any minimum and restoring the initial symmetry.

A further development must be the inclusion of the non-tachyonic modes. Here one may hit the problem  that the considered magnetic string is not a solution of the initial  equations of motion since ${\cal L}_1$, \eqref{3.5} is not zero. As long as the consideration is restricted to the tachyonic modes this is not a problem since ${\cal L}_1$ couples only to the third color component.

\appendix*
\section{Bound state solutions in the flux tube}

In this appendix we demonstrate the solution of the eigenvalue problem \eqref{3.14} and we follow standard methods. The problem can be viewed as a stationary Schr\"odinger equation. It has discrete eigenvalues with $\kappa_l^2>0$ and scattering states  with $\kappa_l^2<0$. The bound state solutions, which correspond to  the tachyonic modes, must decrease for $r \to\infty$.
The solutions can be found in terms of Bessel function in the outside region and in terms of Kummer functions in the inside region,
\eq{3.16}{\phi_l(r)=\frac{1}{N_l}\left(
	\phi_l^{\rm int}(r)\Theta(R-r)
	+\phi_l^{\rm ext}(r)\Theta(r-R)\right),
}
matching  functions and their derivatives by continuity at $r=R$.
This way, for the outside function we make the ansatz
\eq{3.17}{\phi_l^{\rm ext}(r)=
	\beta_lK_\nu(\kappa r)
}
where $\beta_l$ are some constants, $\nu=l-\frac{BR^2}{2}$, $K_\nu(\kappa r)$ is a modified Bessel function and $\kappa$ is still to be found.
To find  the inside function, $\phi_l^{\rm int}(r)$, we follow a well known procedure and make a substitution,
\eq{3.18}{\rho=\frac{B r^2}{2},
}
of the radial variable and the ansatz
\eq{3.19}{\phi_l^{\rm int}(r)=
	\rho^{l/2}e^{-\rho/2}M(\rho).
}
Equation \eqref{3.14} turns into
\eq{3.20}{ \left(
	\rho\pa_\rho+(l+1-\rho)\pa_\rho -\frac{R^2}{2\delta}-l+\frac12\right)M(\rho)=0,
}
where we use the notation $\delta=BR^2/2$, \eqref{3.12}.

The solution  of equation \eqref{3.19}, which is regular at the origin, is the Kummer function, $M(a,l+1,\rho)$,
with the notation
\eq{3.22}{a=\frac{\kappa^2 R^2}{2\delta}+l-\frac12.
}
In order to find the eigenvalues, it is sufficient to match the logarithmic derivatives. For this, we define
\eq{3.23}{ R^{\rm ext}(r)=\kappa \pa_r\ln \phi_l^{\rm ext}(r),
	~~~R^{\rm int}(r)=\kappa \pa_r\ln \phi_l^{\rm int}(r)
}
and demand
\eq{3.24}{R^{\rm ext}(R)=R^{\rm int}(R).
}
It is convenient to introduce another, dimensionless notation, $x=\kappa R$, and to rewrite
\eq{3.25}{R^{\rm ext}(R)&=x\pa_x\ln K_\nu(x),
	\\\nn	R^{\rm int}(R)&=l-\delta+\frac{2\delta a}{l+1}
	\frac{M(a+1,l+2,\delta)}{M(a,l+1,\delta)}
}
where eq. (9.213) from \cite{grad07} was used.

The solutions $\kappa_l$ of equation \eqref{3.24} are the eigenvalues of the operator in the left side of \eqref{3.14}. The coefficients $\beta_l$ and the normalization factors $N_l$ in eq. \eqref{3.16} can be found from matching the functions and from
\eq{3.27}{\int_0^\infty dr\,r\  \phi_l(r )\phi_{l'}(r )=\delta_{ll'},
}
which is the normalization condition.


\begin{thebibliography}{10}
	
	\bibitem{eich11-83-045014}
	Astrid Eichhorn, Holger Gies, and Jan~M. Pawlowski.
	\newblock Gluon condensation and scaling exponents for the propagators in
	yang-mills theory.
	\newblock {\em Phys. Rev. D}, 83:045014, Feb 2011.
	
	\bibitem{savv77-71-133}
	G.K. Savvidy.
	\newblock Infrared instability of vacuum state of gauge theories and asymptotic
	freedom.
	\newblock {\em Phys.~Lett.~B}, 71(1):133--134, 1977.
	
	\bibitem{niel78-144-376}
	N.K Nielsen and P.~Olesen.
	\newblock {Unstable Yang-Mills Field Mode}.
	\newblock {\em Nucl.~Phys.~B}, {144}({2-3}):{376--396}, {1978}.
	
	\bibitem{niel79-160-380}
	H.B. Nielsen and P.~Olesen.
	\newblock {A quantum liquid model for the QCD vacuum: Gauge and rotational
		invariance of domained and quantized homogeneous color fields}.
	\newblock {\em Nucl.~Phys.~B}, 160(2):380--396, 1979.
	
	\bibitem{flor83_SLAC-pub-3244}
	Curt~A. Flory.
	\newblock { Covariant Constant Chromomagnetic Fields and Elimination of the One
		Loop Instabilities}.
	\newblock 1983.
	\newblock Preprint, SLAC-PUB3244, 1983.
	
	\bibitem{leut80-96-154}
	H.~Leutwyler.
	\newblock {Vacuum Fluctuations Surrounding Soft Gluon Fields}.
	\newblock {\em Phys. Lett.}, 96B:154--158, 1980.
	
	\bibitem{savv23-990-116187}
	George Savvidy.
	\newblock Stability of yang mills vacuum state.
	\newblock {\em Nuclear Physics B}, 990:116187, 2023.
	
	\bibitem{skal00-576-430}
	Vladimir Skalozub and Michael Bordag.
	\newblock {Color ferromagnetic vacuum state at finite temperature}.
	\newblock {\em Nucl.~Phys.~B}, 576:430--44, 2000.
	
	\bibitem{ditt81-100-415}
	Walter Dittrich and Volker Schanbacher.
	\newblock {The effective QCD lagrangian at finite temperature}.
	\newblock {\em Physics Letters B}, 100(5):415--419, 1981.
	
	\bibitem{verc08-660-432}
	David Vercauteren and Henri Verschelde.
	\newblock Resolving the instability of the savvidy vacuum by dynamical gluon
	mass.
	\newblock {\em Physics Letters B}, 660(4):432--438, 2008.
	
	\bibitem{kond14-89-105013}
	Kei-Ichi Kondo.
	\newblock {Stability of chromomagnetic condensation and mass generation for
		confinement in SU(2) Yang-Mills theory}.
	\newblock {\em Phys. Rev. D}, 89:105013, May 2014.
	
	\bibitem{nede21-103-114021}
	Sergei Nedelko and Vladimir Voronin.
	\newblock Energy-driven disorder in mean field qcd.
	\newblock {\em Phys. Rev. D}, 103:114021, Jun 2021.
	
	\bibitem{diak02-66-096004}
	Dmitri Diakonov and Martin Maul.
	\newblock {Center-vortex solutions of the Yang-Mills effective action in three
		and four dimensions}.
	\newblock {\em Phys. Rev. D}, 66:096004, 2002.
	
	\bibitem{bord03-67-065001}
	M.~Bordag.
	\newblock {Vacuum energy of a color magnetic vortex}.
	\newblock {\em Phys. Rev.}, D67:065001, 2003.
	
	\bibitem{bord23-55-59}
	M.~Bordag.
	\newblock {Tachyon condensation in a chromomagnetic background field and the
		groundstate of QCD}.
	\newblock {\em Eur.~Phys.~J.~A}, 59:55, 2023.
	\newblock arXiv 2207.08711.
	
	\bibitem{bord2207.08711}
	M.~Bordag.
	\newblock {Tachyon condensation in a chromomagnetic background field and the
		groundstate of QCD}.
	\newblock 2022.
	\newblock arXiv 2207.08711, to appear in EPJA.
	
	\bibitem{grad07}
	I.S. Gradshteyn and I.M. Ryzhik.
	\newblock {\em Table of Integrals, Series and Products}.
	\newblock Academic Press, New York, 2007.
	
\end{thebibliography}

\end{document}